\documentclass[12pt]{article}
\usepackage{color,epsfig,citesort,amsmath,fullpage}

\def\myendproof{{\hfill \vbox{\hrule\hbox{%
   \vrule height1.3ex\hskip0.8ex\vrule}\hrule }}\par}
\newtheorem{theorem}{Theorem}
\newtheorem{lemma}[theorem]{Lemma}
\newtheorem{corollary}[theorem]{Corollary}

\newenvironment{proof}{{\it Proof. }}{\myendproof}

\newcommand{\setof}[1]{\{{#1}\}}
\newcommand{\Xomit}[1]{}

\title{{\bf Locally connected spanning trees on graphs}}
\author{Ching-Chi Lin\thanks{Department of Computer Science and
                              Information Engineering,
                              National Taiwan University,
                              Taipei 10617, Taiwan.}
                     \thanks{Email: d91018@csie.ntu.edu.tw.}
                     \thanks{Institute of Information Science,
                             Academia Sinica,
                             Nankang, Taipei 11507, Taiwan.}
        \and
        Gerard J. Chang\thanks{Department of Mathematics,
                               National Taiwan University,
                               Taipei 10617, Taiwan.
                               Email: gjchang@math. ntu.edu.tw.
                               Supported in part by the National Science
                               Council under grant NSC93-2213-E-002-028.
                               Member of Mathematics Division,
                               National Center for Theoretical Sciences at Taipei.}
        \and
        Gen-Huey Chen$^*$\thanks{Email: ghchen@csie.ntu.edu.tw.}}
\date{\today}
\begin{document}
\maketitle

\begin{abstract}
A locally connected spanning tree of a graph $G$ is a spanning
tree $T$ of $G$ such that the set of all neighbors of $v$ in $T$
induces a connected subgraph of $G$ for every $v\in V(G)$. The
purpose of this paper is to give linear-time algorithms for
finding locally connected spanning trees on strongly chordal
graphs and proper circular-arc graphs, respectively.

\bigskip

\noindent \textbf{Keywords:} algorithm, circular-arc graph,
directed path graph, interval graph, locally connected spanning
tree, proper circular-arc graph, strongly chordal graph.
\end{abstract}

\newpage

\def\skippt{23pt}
\baselineskip \skippt

\section{Introduction}\label{section:intro}

Communication networks or power transmission networks are often
modelled as graphs $G$. The vertices in $V(G)$ represent sites in
the network and the edges in $E(G)$ represent communication lines
or power transmission lines. When delivering a message to a remote
site on a communication network, it is transferred by a path
consisting of many communication lines. Similarly, power
transmission between source site and destination site is
accomplished by a serial of power transmission lines. It is
inexpensive to construct such networks as tree networks. However,
one single site failure would influence the whole network. In
order to guarantee the quality of service,
Farley~\cite{Farley81,Farley82} proposes {\em isolated failure
immune} (IFI) networks.

\begin{figure}[htb]
\centerline{\input{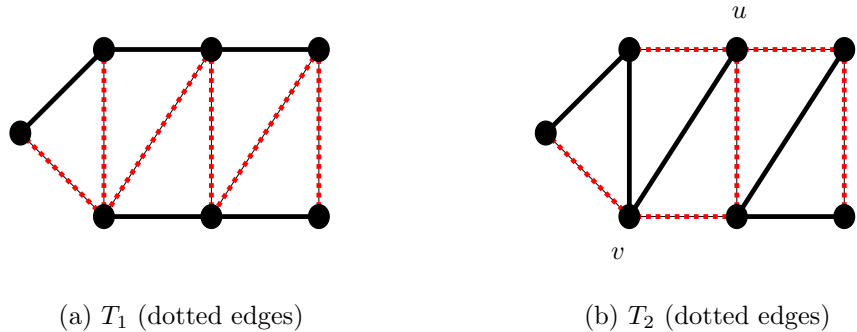}} \caption{Graph $G$ with spanning
trees $T_1$ and $T_2$.} \label{fig:examples}
\end{figure}

Two site failures are {\em isolated} if the sites are not
adjacent. A network is {\em immune} to a set of failures if
transmission between operative sites can be completed under such
failures. A graph is a {\em $2$-tree} if it is either a 2-clique
or it can be produced by adding a new vertex $v$ and two edges
$vx$ and $vy$ to a $2$-tree such that $xy$ is an edge of the
2-tree. It has been shown that an IFI network is minimum if and
only if it is a $2$-tree~\cite{Farley81,Wald83}.
Cai~\cite{Cai97,Cai03} introduced the concept of locally connected
spanning tree and showed that a network containing a locally
connected spanning tree is an IFI network. A {\em locally
connected spanning tree} of a graph $G$ is a spanning tree $T$
such that the set of all neighbors of $v$ in $T$ induce a
connected subgraph of $G$ for every $v\in V(G)$.
Figure~\ref{fig:examples} shows a graph $G$ with a locally
connected spanning tree $T_1$ and a non-locally connected spanning
tree $T_2$. Notice that the set of all neighbors of $u$
(respectively, $v$) in $T_2$ induces a disconnected subgraph in
$G$.

Cai~\cite{Cai03} proved that determining whether a graph contains
a locally connected spanning tree is NP-complete for planar graphs
and split graphs. Furthermore, he also gave a linear-time
algorithm for finding a locally connected spanning tree of a
directed path graph, and a linear-time algorithm for adding fewest
edges to a graph to make a given spanning tree of the graph a
locally connected spanning tree of the augmented graph. Since
split graphs are chordal, determining whether a graph contains a
locally connected spanning tree is NP-complete for chordal graphs.
It is well known that the family of strongly chordal graphs is a
proper subfamily of the family of chordal graphs, and is a proper
superfamily of the family of directed path graphs. In this paper,
we give linear-time algorithms for finding locally connected
spanning trees on strongly chordal graphs and proper circular-arc
graphs, respectively. The former answers an open problem proposed
by Cai~\cite{Cai03}.

The remainder of the paper is organized as follows.
Section~\ref{section:algo-str-cho} describes and analyzes our
algorithm for strongly chordal graphs.
Section~\ref{section:algo-pro-cir} describes and analyzes our
algorithm for proper circular-arc graphs.
Section~\ref{section:conclusion} concludes the paper with an open
question.

We conclude this section at the following two lemmas which are
useful in this paper.
A {\em separating set} $S$ of a graph $G$ is a set $S \subseteq
V(G)$ such that $G-S$ has more than one component. A {\it
cut-vertex} is a vertex that forms a separating set.  A graph is
{\em $k$-connected} if it contains no separating set of size less
than $k$. For $S \subseteq V(G)$, the {\em subgraph induced by}
$S$ is the graph $G[S]$ whose vertex set is $S$ and edge set
$\{xy\in E(G): x,y\in S\}$.

\begin{lemma}[\cite{Cai03}]
\label{lemma:separating-set} If $G$ has a locally connected
spanning tree $T$ and $S$ is a separating set of $G$, then $G[S]$
contains at least one edge of $T$.  Consequently, a graph having a
locally connected spanning tree is $2$-connected.
\end{lemma}

\begin{lemma}\label{lemma:common-neighbor}
Suppose $\setof{x,y}$ is a separating set of $G$ and $H$ is a
component of $G-\setof{x,y}$. If $H$ contains no common neighbor
of $x$ and $y$, then $G$ has no locally connected spanning tree.
\end{lemma}
\begin{proof}
Suppose to the contrary that $G$ has a locally connected spanning
tree $T$. Then there exists a vertex of $H$ connecting $x$ or $y$,
say $x$, through an edge in $T$. Notice that $\setof{x,y}$ is a
separating set of $G$, so $T$ contains the edge $xy$. Since the
neighborhood of $x$ in $T$ induces a connected subgraph in $G$,
$y$ is connected to $H[N_T(x)]$, which implies $x$ and $y$ have a
common neighbor in $H$, a contradiction.
\end{proof}

\section{Algorithm for strongly chordal graphs
                              \label{section:algo-str-cho}}

This section establish a linear-time algorithm for determining
whether a strong chordal graph has a locally connected spanning
tree, and producing one if the answer is positive.  First, some
preliminaries on strongly chordal graphs.

A graph is {\em chordal} (or {\em triangulated}) if every cycle of
length greater than three has a chord, which is an edge joining
two noncontiguous vertices in the cycle. The {\em neighborhood}
$N_G(v)$ of a vertex $v$ is the set of all vertices adjacent to
$v$ in $G$; and the {\em closed neighborhood} $N_G[v]=N_G(v) \cup
\{v\}$. A vertex $v$ is {\em simplicial} if $N_G[v]$ is a clique.
For any ordering $(v_1, v_2, \ldots, v_n)$ of $V(G)$, let $G_i$
denote the subgraph of $G$ induced by $\{v_i, v_{i+1}, \ldots,
v_n\}$. It is well known \cite{Golumbic80} that a graph $G$ is
chordal if and only if it has a {\em perfect elimination order}
which is an ordering $(v_1,v_2,\ldots,v_n)$ of $V(G)$ such that
each $v_i$ is a simplicial vertex of $G_i$.

A {\em strongly chordal} graph is a chordal graph such that every
cycle of even length at least six has a chord that divides the
cycle into two odd length paths.
Farber~\cite{Farber83} proved that a graph is strongly chordal if
and only if it has a {\em strong elimination order} which is an
ordering $(v_1, v_2, \ldots, v_n)$ of $V(G)$ such that
$N_{G_i}[v_j] \subseteq N_{G_i}[v_k]$ for $i \le j \le k$ and
$v_j, v_k \in N_{G_i}[v_i]$.  Notice that a strong elimination
order is also a perfect elimination order.  Anstee and
Farber~\cite{Farber84} presented an $O(n^3)$-time algorithm,
Hoffman, Kolen, and Sakarovitch~\cite{Hoffman85} presented an
$O(n^3)$-time algorithm, Lubiw~\cite{Lubiw87} presented an $O(m
\log^2 m)$-time algorithm, Paige and Tarjan~\cite{Paige87}
presented an $O(m \log m)$-time algorithm and
Spinrad~\cite{Spinrad93} presented an $O(n^2)$-time algorithm for
finding a strong elimination order of a strongly chordal graph
of $n$ vertices and $m$ edges.

According to Lemma \ref{lemma:separating-set}, for a graph to have
a locally connected spanning tree it is necessary that the graph
is $2$-connected.  We now give a necessary and sufficient
condition for a chordal graph, and hence strongly chordal graph,
to be $k$-connected.

\begin{lemma} \label{lemma:k-connected}
Suppose $\sigma=(v_1, v_2, \ldots, v_n)$ is a perfect elimination
order of a chordal graph $G$ and $k<n$ is a positive integer.
Then, $G$ is $k$-connected if and only if $|N_{G_i} (v_i)| \ge k$
for $1 \le i \le n-k$.
\end{lemma}
\begin{proof}
Suppose $P = (v_{i_1}, v_{i_2}, \ldots, v_{i_r})$ is a shortest
$v_i$-$v_n$ path.  We claim that $i=i_1 < i_2 < \ldots < i_r = n$.
Assume to the contrary that $i_{s-1} > i_s$ for some $2 \le s \le
r$.  We may choose $s$ to be as large as possible. As $i_r = n \ge
i_{s-1} > i_s$, we have that $s\le r-1$. By the choice of $s$ we
also have $i_{s+1} > i_s$. Since $v_s$ is a simplicial vertex of
$G_s$, we have $v_{i_{s-1}}, v_{i_{s+1}} \in N(v_{i_s})$ implying
$v_{s-1} v_{s+1} \in E(G)$, contradicting that $P$ is a shortest
path.

($\Rightarrow$) Suppose $G$ is $k$-connected, but $|N_{G_i} (v_i)|
< k$ for some $1 \le i \le n-k$. Then $\sigma'$ obtained from
$\sigma$ by deleting all vertices in $N_{G_i} (v_i)$ is a
simplicial elimination order of $G'=G-N_{G_i} (v_i)$ which is
connected. By the claim above, there is a shortest $v_i$-$v_{n'}$
path $(v_{i_1}, v_{i_2}, \ldots, v_{i_r})$ in $G'$ with $i=i_1 <
i_2 < \ldots < i_r = n'$. This is impossible as $v_{i_2}$ is a
neighbor of $v_i$ in $G-N_{G_i} (v_i)$.

($\Leftarrow$) Suppose $|N_{G_i} (v_i)| \ge k$ for $1 \le i \le
n-k$.  For any subset $S \subseteq V(G)$ of size less than $k$,
let $v_{n'}$ be the vertex of $V(G)-S$ with largest index. Then,
any vertex $v_i$ of $G-S$ other than $v_{n'}$ should have at least
one neighbor $v_{i^*}$ not in $S$ and $i<i^*$. Consequently, every
vertex $v_i$ in $G-S$ has a path connecting to $v_{n'}$.
Therefore, $G-S$ is connected.  This gives the $k$-connectivity of
$G$.
\end{proof}

\bigskip

Now, suppose $(v_1, v_2, \ldots, v_n)$ is a strong elimination
order of $G$. For any neighbor $v_j$ of $v_i$ with $j > i$, let
$\ell(v_i,v_j)=k$ be the minimum index such that $v_k\in
N_G[v_i]\cap N_G[v_j]$. Notice that $\ell(v_i,v_j)$ always exists.
For the case when $v_i$ and $v_j$ has no common neighbors with
indices smaller than $i$ we have $\ell(v_i,v_j)=i$. The {\em
closest neighbor} of a vertex $v_i$ is the vertex $v_{i^*} \in
N_{G_i}(v_i)$ such that $\ell(v_i,v_{i^*}) \le \ell(v_i,v_j)$ for
all $v_j \in N_{G_i}(v_i)$, while tie breaks by choosing  $i^*$
minimum.

In the following, we give an algorithm to determine whether a
strongly chordal graph has a locally connected spanning tree, and
to produce one when the answer is positive. The algorithm first
choose $v_{n-1}v_n$ as an edge of the desired tree.  It then
iterates for $i$ from $n-2$ back to $1$ by adding the edge
$v_iv_{i^*}$ into the tree.  To ensure the $2$-connectivity of the
graph $G$, according to Lemma \ref{lemma:k-connected}, we check if
$|N_{G_i}(v_i)| \ge 2$. When the answer is negative, the graph is
not $2$-connected and so has no locally connected spanning tree.

\bigskip

\baselineskip 15pt
\noindent {\textbf{Algorithm} Strongly-Chordal.}

\noindent {\textbf{Input:} A strongly chordal graph $G$ of order
$n \ge 3$ with a strong elimination order $(v_1, v_2, \ldots,
v_n)$.

\noindent \textbf{Output:} A locally connected spanning tree $T_1$
of $G$ if it exists, and ``NO" otherwise.

\begin{enumerate}
\item For $i=1$ to $n$ do

      ~ ~ ~ Sort $N_G(v_i)$ into
      $v_{i_1}, v_{i_2}, \ldots, v_{i_{d_i}}$,
            where $i_1 < i_2 < \dots < i < i_{p_i} < \dots < i_{d_i}$.

\item For $i=1$ to $n$ do $v_{i^*}=0$.

\item For $j=1$ to $n$ do

      ~ ~ ~ If $v_{j^*} = 0$, then $v_{j^*} = v_{j_{p_j}}$.

      ~ ~ ~ For $k=p_j$ to $d_j-1$ do

      ~ ~ ~ ~ ~ ~ If $v_{(j_k)^*}=0$, then $v_{(j_k)^*}=v_{j_{k+1}}$.

\item If $v_{n-1}$ is adjacent to $v_n$, then let
      $T_{n-1}=v_{n-1}v_n$, else return ``NO".

\item For $i=n-2$ to $1$ step by $-1$ do

      ~ ~ ~ If $|N_{G_i}(v_i)| \le 1$, then return ``NO",
      else let $T_i=T_{i+1}+v_iv_{i^*}$.

\item Return $T_1$.
\end{enumerate}
\baselineskip \skippt

Notice that we may use ``$T_{n-1}=v_{n-1}v_n$" in step 4 of the
algorithm, as $|N_{G_{n-2}}(v_{n-2})| \ge 2$ implying that
$v_{n-1}$ is adjacent to $v_n$. Also, $v_{(n-1)^*}=v_n$, and so we
can interpret step 4 as ``$T_n=\phi$ and $T_{n-1} = T_n + v_{n-1}
v_{(n-1)^*}$''.

\begin{theorem}
\label{theorem:strongly-chordal} For a strongly chordal graph $G$
with a strong elimination order provided, Algorithm
Strongly-Chordal determines in linear-time whether $G$ has a
locally connected spanning tree, and produces one if the answer is
positive.
\end{theorem}
\begin{proof}
We first claim that steps 1 to 3 give the closest neighbor
$v_{i^*}$ of each $v_i$.  Notice that step 1 sorts the neighbors
of each vertex first.   For the case when there are no $i_r < i <
i_s$ with $v_{i_r}$ adjacent to $v_{i_s}$, the closest neighbor
$v_{i^*}$ is the neighbor of $v_i$ of minimum index which is
larger than $i$, namely $v_{i_{p_i}}$. For the other case,
$v_{i^*}$ is obtained by finding a minimum index $j$ such that
$i=j_k$ and $i^*=j_{k+1}$ for some $k$ with $p_j \le k \le d_j-1.$
These are taken care of in steps 2 and 3.

When the algorithm returns a ``NO", according to Lemmas
\ref{lemma:k-connected} and \ref{lemma:separating-set}, the graph
$G$ has no locally connected spanning tree.  We now assume that
the algorithm returns $T_1$.  In this case, $|N_{G_i}(v_i)| \ge 2$
for all $i\le n-2$.  By Lemma \ref{lemma:k-connected}, $G$ is
$2$-connected. We first claim that $T_{i+1}$ has an edge $v_j
v_{i^*}$ whose end vertices are neighbors of $v_i$ for $i \le
n-2$.

When $v_{i^*}=v_{i_{d_i}}$, let $v_j = v_{i_{d_i-1}}$.  Since
$v_j$ and $v_{i^*}$ are neighbors of $v_i$ with $i < j < i^*$, we
have $v_j v_{i^*}\in E(G)$ and $\ell(v_j,v_{i^*})\le i$. If
$v_{j^*}=v_{i^*}$, then $T_{i+1}$ has an edge $v_j v_{i^*}=v_j
v_{j^*}$ whose end vertices are neighbors of $v_i$ as desired.
Now, suppose $v_{j^*} \ne v_{i^*}$. By the choice of $v_j$, we
have that $v_{j^*}$ is not a neighbor of $v_i$.  And so
$\ell(v_j,v_{j^*}) < \ell(v_j,v_{i^*})\le i$.  Let $k =
\ell(v_j,v_{j^*})$. Then $v_i \in N_{G_k}[v_j] \subseteq
N_{G_k}[v_{j^*}]$, which violates that $v_{j^*}$ is not a neighbor
of $v_i$.

When $v_{i^*}=v_{i_r}$ with $r<d_i$, consider the closest neighbor
$v_{i^{**}}$ of $v_{i^*}$. Let $k=\ell(v_{i^{*}},v_{i^{**}})$.
Then, we have $k=\ell(v_{i^{*}},v_{i^{**}}) =
\ell(v_{i_r},v_{(i_r)^*})\le \ell(v_{i_r},v_{i_{d_i}}) \le i$.  It
follows that $v_i \in N_{G_k}[v_{i^*}] \subseteq
N_{G_k}[v_{i^{**}}]$, and so $T_{i+1}$ has an edge $ v_{i^*}
v_{i^{**}}$ whose end vertices are neighbors of $v_i$ as desired.

We shall prove that $T_i$ is a locally connected spanning tree of
$G_i$ for each $i$ by induction on $i$ from $n$ back to $1$. The
assertion is clearly true for $i\ge n-1$. Suppose $T_{i+1}$ is a
locally connected spanning tree of $G_{i+1}$.  To see $T_i$ is a
locally connected spanning tree of $G_i$, we only need to verify
that $G_i[N_{T_i}(v_i)]$ and $G_i[N_{T_i}(v_{i^*})]$ are connected
as $T_i=T_{i+1}+v_iv_{i^*}$. Since $v_i$ is a leaf in $T_i$,
$G_i[N_{T_i}(v_i)]$ is connected. According to the facts that
$G_{i+1}[N_{T_{i+1}}(v_{i^*})]$ is connected and that $T_{i+1}$
has an edge connecting $v_{i^*}$ and a neighbor of $v_i$ in
$G_{i+1}$ for $i \le n-2$, it follows that $G_i[N_{T_i}(v_{i^*})]$
is connected.

These prove the correctness of the algorithm.

We finally argue that the time complexity for the algorithm is
linear. In step $1$, we may sort the neighbors of each $v_i$ by
adding $v_i$ into the adjacent list of each neighbor of $v_i$ from
$i=1$ to $n$.  So totally, step $1$ takes $O(n+m)$ time. It is
easy to see that the other steps also take linear time.
\end{proof}

\begin{corollary}\label{corollary:strongly-chordal}
If $G$ is a strongly chordal graph, then $G$ has a locally
connected spanning tree $T$ if and only if it is $2$-connected.
\end{corollary}

\section{Algorithm for proper circular-arc graphs
                              \label{section:algo-pro-cir}}

This section establishes a linear-time algorithm for determining
whether a proper circular-arc graph has a locally connected
spanning tree, and producing one if the answer is positive. First,
some preliminaries on circular-arc graphs.

A {\em circular-arc graph} $G$ is the intersection graph of a
family $F$ of arcs in a circle, with vertices of $G$ corresponding
to arcs in $F$ and two vertices in $G$ are adjacent if and only if
their corresponding arcs in $F$ overlap.
McConnell~\cite{McConnell01,McConnell03} gave a linear-time
algorithm to recognize circular-arc graphs. As a byproduct, an
intersection model $F$ of circular-arc graph $G$ can be
constructed in linear time. A family $F$ is said to be {\em
proper} if no arc in $F$ is contained in another.

For a vertex $v$ of $G$, let $a(v)$ denote the corresponding arc
in $F$. An arc $a(v)$ that begins at endpoint $h(v)$ and ends at
endpoint $t(v)$ in a counterclockwise traversal is denoted by
$[h(v),t(v)]$, where $h(v)$ is the {\em head} of $a(v)$ and $t(v)$
is the {\em tail} of $a(v)$. Assume without loss of generality
that all arc endpoints are distinct and no arc covers the entire
circle. A {\em segment} $(s,t)$ of a circle is the continuous part
that begins at endpoint $s$ and ends at endpoint $t$ in a
counterclockwise traversal. The segment $(s,t)$ is considered as
not containing points $s$ and $t$ and segment $[s,t]$ is
considered as containing $s$ and $t$. Similarly, $[s,t)$ and
$(s,t]$ are segments containing $s$ but not $t$; and not
containing $s$ but $t$. The {\em density} $d(v)$ of the arc $a(v)$
is the number of arcs, including $a(v)$, in $F$ that contain
$h(v)$.

First, a lemma on circular-arc graphs.

\begin{lemma}\label{lemma:density}
If a circular-arc graph $G$ has at least four corresponding arcs
with $d(v)\le 2$ in $F$, then $G$ has no locally connected
spanning tree.
\end{lemma}
\begin{proof}
Suppose $a(v_p)$, $a(v_q)$, $a(v_r)$ and $a(v_s)$ are four arcs of
$F$ with density at most $2$ in a counterclockwise traversal, see
Figure \ref{fig:density}. Let $a(v_{p'})$, $a(v_{q'})$,
$a(v_{r'})$ and $a(v_{s'})$ be the arcs which contain the heads of
$a(v_p)$, $a(v_q)$, $a(v_r)$ and $a(v_s)$, respectively. We assume
that $a(v_{p'})$, $a(v_{q'})$, $a(v_{r'})$ or $a(v_{s'})$ is empty
when $d(v_p)$, $d(v_q)$, $d(v_r)$ or $d(v_s)$ is $1$,
respectively.

If $a(v_{p'})$ exists and contains $h(v_s)$, i.e.
$a(v_{p'})=a(v_{s'})$, then $v_{p'}$ is a cut-vertex of $G$ since
no arc in $F-\{a(v_{p'})\}$ crosses the points $h(v_p)$ and
$h(v_s)$.  In this case, $G$ has no locally connected spanning
tree.  So, we may assume that $a(v_{p'})$ does not cross $h(v_s)$.
Similarly, we may assume that $a(v_{p'})$ does not cross $h(v_q)$.
Then, $a(v_{p'})$ is contained in $[h(v_s),h(v_q))$. Similarly, if
$a(v_{r'})$ exists then it is contained in $[h(v_q),h(v_s))$.
Therefore, $a(v_{p'})$ and $a(v_{r'})$ do not overlap, which
implies that $v_{p'}v_{r'}$ is not an edge of $G$. Again,
$\setof{v_{p'},v_{r'}}$ is a separating set of $G$, since no arc
in $F-\{a(v_{p'}),a(v_{r'})\}$ crosses the points $h(v_p)$ and
$h(v_r)$. According to Lemma \ref{lemma:separating-set}, $G$ has
no locally connected spanning tree as $v_{p'}v_{r'} \not \in
E(G)$. Notice that the proof covers the case when $a(v_{p'})$ or
$a(v_{r'})$ is empty.
\end{proof}

\begin{figure}[htb]
\centerline{\input{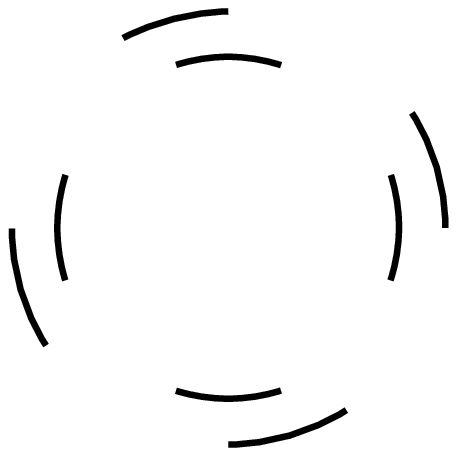}} \caption{Graph $G$ with four
corresponding arcs of density at most $2$.} \label{fig:density}
\end{figure}

Suppose $G$ is a circular-arc graph in which $d(v)=1$ for some
vertex $v$, then $G$ is in fact an interval graph. In this case,
results in the previous section can be used to determine whether
$G$ has a locally connected spanning tree as interval graphs are
strongly chordal.  Notice that the ordering from left to right of
the right endpoints of the intervals in an interval representation
of $G$ is a strong elimination order. Therefore, without lost of
generality, {\it we may assume that $d(v)\ge 2$ in $F$ for each
vertex $v$ of $G$}.

We now turn attention to the algorithm for finding locally
connected spanning trees on proper circular-arc graphs. In this
case, if we identify $a(v_1)$, then let $(a(v_2), a(v_3), \ldots,
a(v_n))$ be the ordering of arcs in $F-\{a(v_1)\}$ such that
$h(v_i)$ is encountered before $h(v_j)$ in a counterclockwise
traversal from $h(v_1)$ if $i<j$.  Since $d(v) \ge 2$ for all
vertices $v$, it is the case that $G$ is $2$-connected as $(v_1,
v_2, \ldots, v_n, v_1)$ is a Hamiltonian cycle.

By Lemma~\ref{lemma:density}, if $G$ has at least four
corresponding arcs with $d(v) = 2$ in $F$, then $G$ has no locally
connected spanning tree. Therefore, we only need to treat the case
when $G$ has at most three corresponding arcs in $F$ with density
equal to $2$. We divide the problem into three cases. For the case
when $G$ has at most one corresponding arc in $F$ with density
equal to $2$, the algorithm is similar to the one for interval
graphs. For the cases $G$ has two or three, the graph $G$ has
special structures, we design algorithm by using these properties. \\

\baselineskip 8pt
\noindent {\textbf{Algorithm} Proper-Circular-Arc.}

\noindent {\textbf{Input:} A proper circular-arc graph $G$ of
order $n \ge 3$ and with an intersection model $F$ such that $d(v)
\ge 2$ for all vertices $v$.

\noindent \textbf{Output:} A locally connected spanning tree $T$
of $G$, if it exists, and ``NO" otherwise.

\begin{enumerate}
\item If $G$ has at least four corresponding arcs in $F$ with
density equal to $2$, then return ``NO''.

\item If $G$ has at most one corresponding arc in $F$ with density
equal to $2$.

\begin{enumerate}
\item If $G$ has one corresponding arc in $F$ with density equal
to $2$, then let $a(v_1)$ contain its head in $F$. Otherwise, let
$a(v_1)$ be an arbitrary arc in $F$.

\item Let $T_1=\phi$. Let $T_i=T_{i-1}+v_iv_{i-1}$ for $i=2$ to
$n$.
\end{enumerate}

\item If $G$ has exactly two corresponding arcs in $F$ with
density equal to $2$.

\begin{enumerate}
\item If the two arcs overlap. Let $a(v_1)$ and $a(v_2)$ be the
two arcs in $F$ such that $a(v_1)$ contains the head of $a(v_2)$.
Do step $2(b)$.

\item Otherwise, suppose $a(v_1)$ and $a(v_k)$ contain the heads
of the two arcs, respectively. We may assume that $a(v_1)$
contains the head of $a(v_k)$, if they overlap. If $v_1$ and $v_k$
have no common neighbor $v_z$ with $z>k$ or $v_1v_k\not \in E(G)$,
then return ``NO''. Otherwise, let $T=\setof{v_1v_i|~i=2,\ldots,k
\text{ or }  i=z} \cup \setof{v_zv_i|~i=k+1,\ldots,n \text{ but }
i\not=z}$.
\end{enumerate}

\item If $G$ has exactly three corresponding arcs in $F$ with
density equal to $2$.

\begin{enumerate}
\item  Suppose $a(v_1)$, $a(v_p)$ and $a(v_q)$ contain the heads
of the three arcs in a counterclockwise traversal, respectively.

\item If all of $\setof{v_1,v_p}$, $\setof{v_1,v_q}$ and
$\setof{v_p,v_q}$ are separating sets of $G$ or one of the three
edges $v_1v_p$, $v_1v_q$ and $v_pv_q$ does not belong to $E(G)$,
then return ``NO''.  Otherwise, we may assume that
$\setof{v_p,v_q}$ is not a separating set of $G$.

\item Let $T=\setof{v_1v_i|~i=2,\ldots,n}$.
\end{enumerate}

\item Return $T$.
\end{enumerate}
\baselineskip 24pt

\begin{figure}[htb]
\centerline{\input{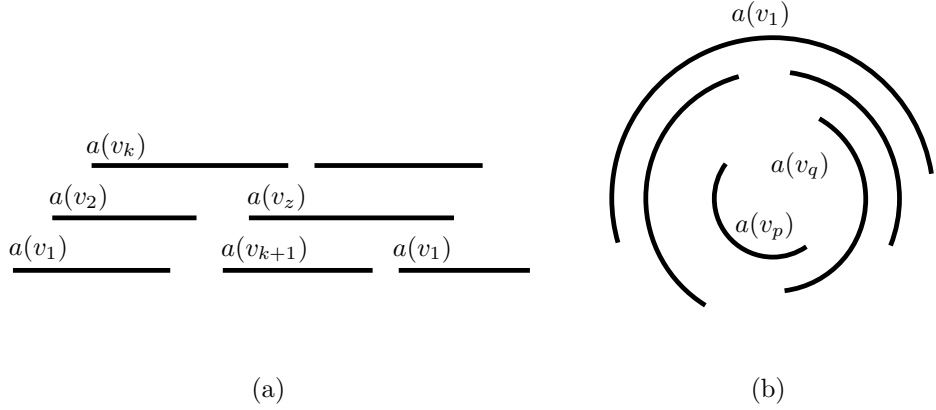}} \caption{Two examples for
Algorithm Proper-Circular-Arc. (a) $G$ contains exactly two
corresponding arcs in $F$ with density equal to $2$. (b) $G$
contains exactly three corresponding arcs in $F$ with density
equal to $2$.} \label{fig:2-and-3-node}
\end{figure}

Now, we prove the correctness of the above algorithm. If the
algorithm outputs a tree $T$, we verify that it is a locally
connected spanning tree. Otherwise, we show that $G$ lacks one
necessary condition of having a locally connected spanning tree.
In the following, let $G_i = G[\{v_1, v_2, \ldots, v_i\}]$.

\begin{lemma} \label{lemma:proper-circular-correctness}
Algorithm Proper-Circular-Arc outputs a locally connected spanning
tree $T$ of $G$, if it exists.
\end{lemma}
\begin{proof} By Lemma~\ref{lemma:density}, if $G$ has at least four
corresponding arcs with $d(v) = 2$ in $F$, then $G$ has no locally
connected spanning tree. Therefore, we need only to consider the
cases when $G$ has at most three corresponding arcs with $d(v) =
2$ in $F$.

We first consider the case when $G$ has at most one corresponding
arc in $F$ with density equal to $2$. Notice that if $v_i$ has $k$
neighbors in $G_i$, then the neighbors of $v_i$ are
$v_{i-1}\ldots,v_{i-k}$. Since $d(v_i)\ge 3$, the vertex $v_i$ has
at least $2$ neighbors $v_{i-1}$ and $v_{i-2}$ in $G_i$ for $i \ge
3$. We shall prove that $T_i$ is a locally connected spanning tree
of $G_i$ by induction on $i$. The claim is true for $T_2$. By the
induction hypothesis, $T_{i-1}$ is a locally connected spanning
tree of $G_{i-1}$ and so $G_{i-1}[N_{T_{i-1}}(v_{i-1})]$ is
connected. We know that $T_{i-1}$ contains an edge connecting
$v_{i-1}$ and $v_{i-2}$ in $G_{i-1}$ for $i \ge 3$ as we have
$T_{i-1}=T_{i-2}+v_{i-1}v_{i-2}$. Since $v_{i-1}$ and $v_{i-2}$
are neighbors of $v_i$, $G_i[N_{T_i}(v_{i-1})]$ and
$G_i[N_{T_i}(v_i)]$ are connected, $T_i$ is a locally connected
spanning tree of $G_i$.

Next, consider the case when $G$ has two corresponding arcs in $F$
with density equal to $2$. If the two arcs overlap, we have
$d(v_1)=d(v_2)=2$ and $d(v_i)\ge 3$ for $i \ge 3$. The remainder
of the proof for this case is similar to the above case.
Otherwise, we have $d(v_2)=d(v_{k+1})=2$ and the two vertices
$v_2$ and $v_{k+1}$ are in different component of
$G-\setof{v_1,v_k}$. See Figure~\ref{fig:2-and-3-node}(a) for an
example. By Lemma~\ref{lemma:common-neighbor}, if $v_1$ and $v_k$
do not have common neighbor $v_z$ with $z>k$, then $G$ has no
locally connected spanning tree. For the case when such $v_z$
exists, we prove that the output $T$ is a locally connected
spanning tree in this case by showing that each edge in $T$ is
also in $E(G)$ and the two components $G[N_T(v_1)]$ and
$G[N_T(v_z)]$ are connected. Consider $v_1v_k\in E(G)$. Since
$a(v_1)$ contains the head of $a(v_k)$, the arc $a(v_i)$ would
also contain $h(v_k)$ for $1<i<k$. It follows that $v_iv_1, v_iv_k
\in E(G)$ for $1<i<k$. Notice that $v_z$ is a common neighbor of
$v_1$ and $v_k$ with $z > k$. Therefore, each vertex in $N_T(v_1)$
is adjacent to $v_1$ in $G$ and $G[N_T(v_1)]$ is connected. Since
$a(v_z)$ contains the tail of $a(v_k)$ and the head of $a(v_1)$,
$a(v_i)$ would also contain $t(v_k)$ for $k < i < z$ and $a(v_i)$
would also contain $h(v_1)$ for $z < i \le n$. It follows each
vertex in $N_T(v_z)$ is also adjacent to $v_z$ in $G$ and both of
the two components $G[v_{k+1},\ldots,v_{z-1}]$ and
$G[v_{z+1},\ldots,v_{n},v_1]$ are connected. Since the density of
$a(v_{z+1})$ is at least $3$, vertex $v_{z-1}$ is adjacent to
$v_{z+1}$, where $v_{z+1}=v_1$ if $z=n$. Therefore, $G[N_T(v_z)]$
is connected.

Finally, consider the case when $G$ has three corresponding arcs
in $F$ with density equal to $2$. Choose any two vertices of
$\setof{v_1,v_p,v_q}$, if one does not succeed another in a
counterclockwise traversal, the two vertices form a separating set
of $G$. If all of $\setof{v_1,v_p}$, $\setof{v_1,v_q}$ and
$\setof{v_p,v_q}$ are separating sets of $G$, then $T$ should
contain the cycle $(v_1,v_p,v_q)$. Thus, the two vertices is
either a separating set of $G$ or one succeeds another in a
counterclockwise traversal. It follows that the three edges
$v_1v_p$, $v_1v_q$ and $v_pv_q$ should belong to $E(G)$ and at
least one of the three sets is not a separating set, if $G$
contains a locally connected spanning tree. To see the output $T$
is a locally connected spanning tree of $G$ in this case, it
suffices to show that $G-v_1$ is connected and $G$ contains edges
containing $v_1$ and $v_i$ for $ i \ge 2$. Consider the case when
$G$ contains the three edges $v_1v_p$, $v_1v_q$ and $v_pv_q$, and
$\setof{v_p,v_q}$ is not a separating set of $G$, i.e., $q=p+1$.
See Figure~\ref{fig:2-and-3-node}(b) for an example. Since
$a(v_p)$ contains the tail of $a(v_1)$ and $a(v_q)$ contains the
head of $a(v_1)$, $a(v_i)$ contains $t(v_1)$ for $2 \le i \le p$
and $a(v_i)$ contains $h(v_1)$ for $q \le i \le n$. Therefore,
$v_iv_1 \in E(G)$ for $2 \le i \le n$ and the two components
$G[v_2,\ldots,v_p]$ and $G[v_q,\ldots,v_n]$ are connected. Since
$v_p$ and $v_q$ are adjacent, $G-v_1$ is connected.
\end{proof}

\bigskip

Now, we prove the algorithm runs in linear time. Recognizing the
corresponding arcs in $F$ whose density are equal to $2$ and
determining the order $(v_1,v_2,\dots,v_n)$ could be done by
traversal the intersection model $F$ in counterclockwise order,
which takes $O(n)$ time. Therefore, Step $1$ takes linear time.
Notice that, if the arcs with density equal to $2$ are recognized,
constructing a locally connected spanning tree in the three
difference cases all take $O(n)$ time. Step $2$ takes linear time.
Consider the Step $3$. It takes $O(1)$ time to check whether
$a(v_1)$ and $a(v_k)$ are overlap and takes $O(n)$ time to check
whether there exists a common neighbor $v_z$ with $z > k$. Step
$3$ takes $O(n)$ time. It also takes constant time to determine
whether any two vertices of $\setof{v_1,v_p,v_q}$ is a separating
set of $G$ by checking whether one succeeds another in a
counterclockwise traversal. Thus, this algorithm runs in linear
time.

\begin{theorem}
\label{theorem:proper-circular} For a proper circular-arc graph
$G$ with an intersection model $F$ provided, Algorithm
Proper-Circular-Arc determines in linear-time whether $G$ has a
locally connected spanning tree, and produces one if the answer is
positive.
\end{theorem}

\section{Conclusion   \label{section:conclusion}}

In this paper, we present two algorithms for finding locally
connected spanning trees on strongly chordal graphs and proper
circular-arc graphs, respectively. The former answers an open
problem proposed by Cai~\cite{Cai03}.  It is an interesting
problem to design an algorithm for finding locally connected
spanning trees on circular-arc graphs or to prove that it is
NP-complete.

\small
\bibliographystyle{abbrv}
\bibliography{lcst}
\end{document}